\documentclass[prl,twocolumn,superscriptaddress,floatfix,noshowpacs,10pt,longbibliography]{revtex4-1}%
\usepackage{graphicx,bm,times}
\usepackage{amsmath}
\usepackage{amsfonts}
\usepackage{amssymb}
\usepackage{bm}
\usepackage{color}
\usepackage{times}
\usepackage[%
  colorlinks=true,
  urlcolor=blue,
  linkcolor=blue,
  citecolor=blue
  ]{hyperref}

\begin{document}

\title{Quenched nematic criticality separating two superconducting domes \\ in an iron-based superconductor under pressure}

\author{P. Reiss}
\email[corresponding author:]{pascal.reiss@physics.ox.ac.uk}
\affiliation{Clarendon Laboratory, Department of Physics,
University of Oxford, Parks Road, Oxford OX1 3PU, UK}

\author{D. Graf}
\affiliation{National High Magnetic Field Laboratory and Department of Physics, Florida State University, Tallahassee, Florida 32306, USA}

\author{A. A. Haghighirad}
\affiliation{Clarendon Laboratory, Department of Physics,
University of Oxford, Parks Road, Oxford OX1 3PU, UK}
\affiliation{Institut fur Festk\"orperphysik, Karlsruhe Institute of Technology, 76021 Karlsruhe, Germany}

\author{W. Knafo}
\affiliation{Laboratoire National des Champs Magn\'etiques Intenses (LNCMI-EMFL),
UPR 3228, CNRS-UJF-UPS-INSA, 143 Avenue de Rangueil, 31400 Toulouse, France}

\author{L. Drigo}
\affiliation{Laboratoire National des Champs Magn\'etiques Intenses (LNCMI-EMFL),
UPR 3228, CNRS-UJF-UPS-INSA, 143 Avenue de Rangueil, 31400 Toulouse, France}
\affiliation{G\'eosciences Environnement Toulouse (CNRS), 31400 Toulouse, France}

\author{M. Bristow}
\affiliation{Clarendon Laboratory, Department of Physics, University of Oxford, Parks Road, Oxford OX1 3PU, UK}

\author{A. J. Schofield }
\affiliation{School of Physics and Astronomy, University of Birmingham, Edgbaston, Birmingham B15 2TT, UK}

\author{A. I. Coldea}
\email[corresponding author:]{amalia.coldea@physics.ox.ac.uk}
\affiliation{Clarendon Laboratory, Department of Physics, University of Oxford, Parks Road, Oxford OX1 3PU, UK}

\begin{abstract}

The nematic electronic state and its associated nematic critical fluctuations have emerged as
potential candidates for superconducting pairing in various unconventional superconductors. 
However, in most materials their coexistence with other magnetically-ordered phases poses significant challenges in establishing their importance. Here, by combining chemical and hydrostatic physical pressure in FeSe$_{0.89}$S$_{0.11}$, we provide a unique access to a clean nematic quantum phase transition in the absence of a long-range magnetic order.
We find that in the proximity of the nematic phase transition, there is an unusual non-Fermi liquid behavior in resistivity at high temperatures that evolves into a Fermi liquid behaviour at the lowest temperatures. From quantum oscillations in high magnetic fields, we trace the evolution of the Fermi surface and electronic correlations as a function of applied pressure.
We detect experimentally a Lifshitz transition that separates two distinct superconducting regions: one emerging from the nematic electronic phase with a small Fermi surface and strong electronic correlations and the other one with a large Fermi surface and weak correlations
that promotes nesting and stabilization of a magnetically-ordered phase at high pressures. 
The lack of mass divergence suggests that the nematic critical fluctuations are quenched by the strong coupling to the lattice. This establishes that superconductivity is not enhanced at the
nematic quantum phase transition in the absence of magnetic order. 
\end{abstract}
\date{\today}
\maketitle


An electronic nematic ordered state can occur in a strongly interacting electronic system in which a Fermi surface undergoes a spontaneous distortion to a shape with lower symmetry compared to the underlying crystal lattice.
The observation of electronic nematic order in different families of high-temperature superconductors 
imply that the same interactions may be involved in stabilizing both the nematic and superconducting states \cite{Fernandes2013,Lederer2017,Sprau2016}.
However, the presence of other competing electronic phases, such as spin or charge-density waves neighbouring superconductivity
can obscure the relevance of the nematic fluctuations in superconducting pairing.
Hydrostatic pressure is a powerful method not only to fine-tune across different electronic phase transitions
and disentangle competing electronic phases, but also to access new regimes, as often found in organic superconductors
and iron-based superconductors \cite{Furukawa2015,Matsuura2017}.

FeSe is a striking example
of a nematic superconductor  in which applied pressure leads to a four-fold 
increase in its bulk superconductivity (from $ 9$\,K towards a high-critical temperature, $T_c \sim 40$\,K) \cite{Mizuguchi2008,Medvedev2009}.
In the normal state, the nematic phase transition of FeSe is suppressed with increasing pressure, but the quantum phase transition is masked by an emerging magnetic ordering stabilized under high pressures \cite{Terashima2015,Terashima2016,Wang2015,Wang2016,Sun2017,Kothapalli2016,Imai2009,Bendele2010}.
As the nematic and magnetic phases of FeSe are intertwined in the pressure-temperature phase diagram, it is
difficult to establish the roles played by nematic or spin fluctuations for pairing
and stabilizing a high-$T_c$ state.
Similarly to applied external pressure, isoelectronic sulfur substitution in FeSe$_{1-x}$S$_{x}$, equivalent to internal positive chemical pressure, suppresses the nematic order, but does not stabilize a magnetically ordered state \cite{Watson2015a,Watson2015c,Coldea2016,Reiss2017,Hosoi2016}. Consequently, by combining chemical pressure and hydrostatic pressure, 
the nematic quantum critical point is unmasked, as the magnetic
order is shifted to higher pressures with increasing sulfur concentration \cite{Matsuura2017,Hosoi2016,Xiang2017,Yip2017}.
This opens a unique path for studying the nature
of nematic criticality in detail using hydrostatic pressure
as a clean tuning parameter and to probe the role of 
nematic fluctuations in stabilizing superconductivity.

As superconductivity depends not only on the origin of the attractive pairing interaction but
also on the details of the Fermi surface from which it emerges \cite{Coldea2017},
understanding the changes in the electronic structure as a function of different tuning parameters is paramount.
Furthermore, the size of the Fermi energy triggers
different electronic instabilities either a nematic order at small values, superconductivity in an intermediate regime
or a magnetically-ordered state for large energies \cite{Chubukov2016}.
To test these different regimes in FeSe$_{1-x}$S$_{x}$, one requires experimental probes which give access to the Fermi surface under applied pressures.
Quantum oscillations are a powerful technique to access the evolution of the electronic structure and correlations under extreme conditions.
In FeSe, quantum oscillations showed that the extremal areas of the Fermi surface increase in the nematic phase,
and a sudden reconstruction was interpreted to occur in the magnetic phase at higher pressures \cite{Terashima2015,Terashima2016}.
On the other hand, quantum oscillations and ARPES studies using chemical pressure detected a continuous growth of the Fermi surface of FeSe$_{1-x}$S$_{x}$, as the highly in-plane-distorted Fermi surface of FeSe becomes isotropic in the tetragonal phase \cite{Coldea2016,Watson2015c,Reiss2017}.
Additionally, quantum oscillations detect an unusual 
Lifshitz transition in FeSe$_{1-x}$S$_{x}$, 
which coincides with the disappearance of the nematic state \cite{Coldea2016} and raises the prospect that these effects may be relevant for stabilising superconductivity.

In this Article we explore the electronic behaviour across the pressure-temperature phase diagram of FeSe$_{0.89}$S$_{0.11}$, in the
absence of long-range magnetic order. Using external pressure, we finely tune the system 
across it critical nematic region to understand the role played by
the electronic structure, scattering and correlations in stabilizing superconductivity.
We identify different regimes of scattering  in the proximity of the
nematic and the high-$T_c$ phases from detailed resistivity measurements under pressure.
The nematic phase is surrounded by
an extended region with a $\sim T^{3/2}$ resistivity,
different from a prototypical quantum critical behaviour.
Remarkably, at low temperatures a Fermi liquid behaviour
is recovered across the nematic critical region.
Quantum oscillations identify a Lifshitz transition at the nematic quantum phase transition,
that separates two types of superconducting domes
having different Fermi surface sizes and different strength of electronic correlations.
Importantly,  the cyclotron effective masses are significantly suppressed across the nematic phase transition, as well as superconductivity.
We interpret this behaviour as a signature of quenched critical fluctuations due to a strong nematoelastic coupling to the lattice \cite{Paul2017}.

\begin{figure}[htbp]
	\centering
	\includegraphics[trim={0cm 0cm 0cm 0cm}, width=1\linewidth,clip=true]{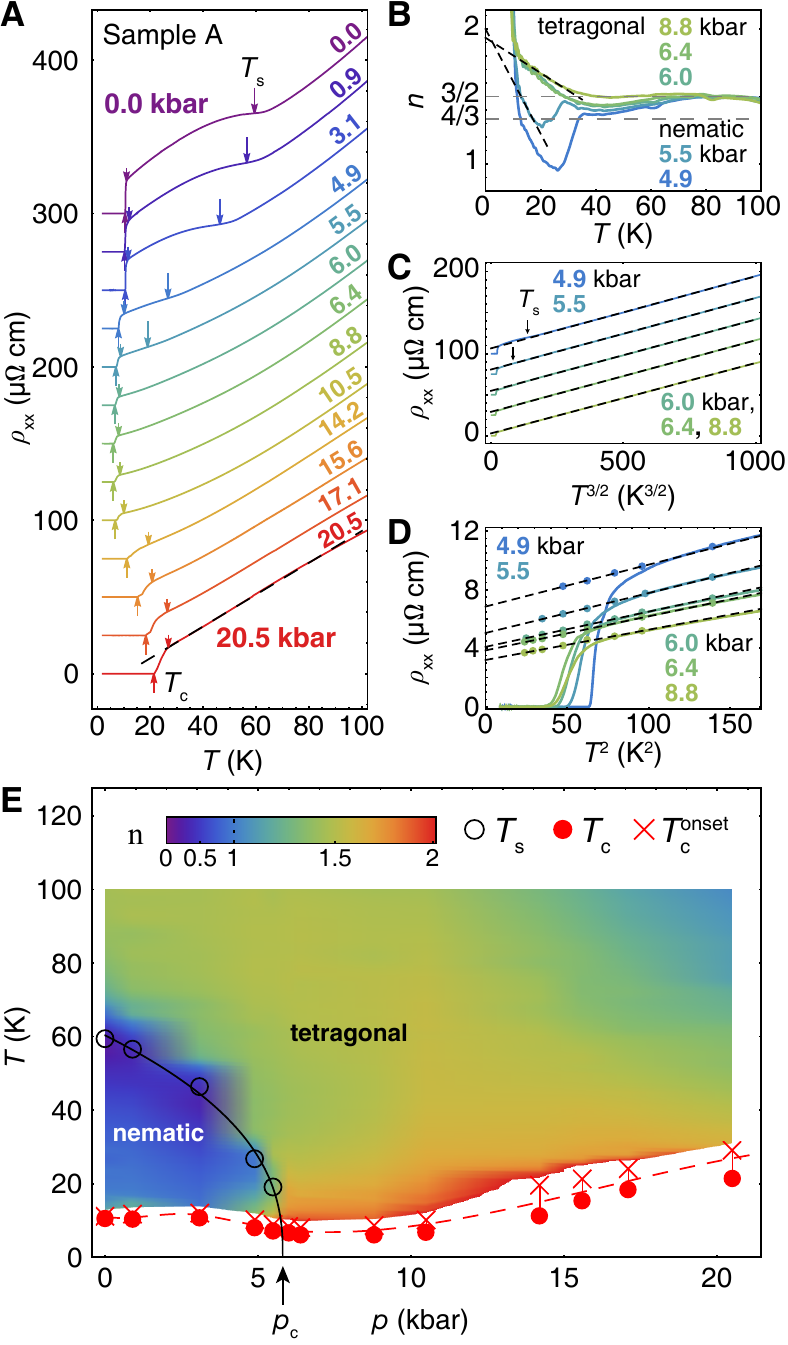}
	\caption{{\bf Transport under pressure in FeSe$_{0.89}$S$_{0.11}$.}
		(\textbf{A}) Temperature dependence of resistivity for different pressures. The transition temperature into the nematic state occurs at  $T_s$ and 
into the superconducting phase at $T_c$. The onset and offset of superconductivity are indicated by crosses and solid circles, respectively in E.
		(\textbf{B}) Local resistivity exponent $n$ for pressures close to $p_c$. Dashed lines are linear fits at low temperatures above the onset of superconductivity.
		(\textbf{C}) High temperature resistivity plotted against $T^{3/2}$. Dashed lines are linear fits. Data in panels \textbf{A} and \textbf{C} are shifted vertically for clarity.
		(\textbf{D}) Low-temperature resistivity (solid lines) and normal state resistivity extrapolated from symmetrized magnetic field measurements (points) plotted against $T^2$ (Fig.S2). Dashed lines are fits to Fermi liquid behaviour, $\rho= \rho_0 +A T^2$,
the slope gives the $A$ coefficient and $\rho_0$ is the zero-temperature residual resistivity. Fit residuals are shown in Fig.~S3.
		(\textbf{E}) Pressure-temperature phase diagram. The color map represents the local resistivity exponent $n$ ($\rho \sim T^n$), see color scale in inset. Symbols indicate the transition temperatures defined in panel (A). The dashed line is a guide to the eye. The solid line is a fit of $T_s \sim (p-p_c)^\epsilon$ giving $\epsilon \sim 0.4$ and $p_c = 5.8$\,kbar.}
	\label{fig1}
\end{figure}

 {\bf Temperature dependence of resistivity with applied pressure.}
Figure~\ref{fig1}A shows the evolution of the temperature dependence
of the resistivity as a function of  applied pressure below $100$\,K
(the full range up to $300$\,K is shown in Fig.~S1).
With increasing pressure, the nematic transition temperature $T_s$ is quickly suppressed from $60$\,K at ambient pressure until it becomes unobservable
around $p_c$ (Fig.~S1). By extrapolating the pressure evolution of $T_s$ to zero, we can locate the position of the nematic quantum phase transition at $p_c\sim5.8(5)$\,kbar in Fig.~\ref{fig1}E.
Within the nematic phase, there is only a weak pressure dependence of the superconducting transition temperature $T_c$ (defined as the zero resistivity temperature) similar to previous findings using chemical pressure \cite{Coldea2016,Reiss2017}.
Outside the nematic phase, superconductivity is suppressed rather than enhanced in the proximity  of the nematic quantum phase transition and reaches a minimum $T_c \sim 6.4$\,K (at $\sim 7$\,kbar) close to $p_c$.
However, superconductivity is enhanced significantly towards $22$\,K at $20$\,kbar, and
$T_c$ was reported to reach $30$\,K  at higher pressures at around $30$\,kbar 
(see Fig.~\ref{fig4}A) \cite{Matsuura2017}.
The superconducting transition at high pressures broadens significantly compared with the low-pressure region (by a factor $\approx 10$),
suggesting that the superconducting phase becomes rather inhomogeneous at high pressures or that it coexists with another electronic phase.
In contrast to FeSe where nematicity, magnetism and superconductivity may coexist under pressure \cite{Terashima2015,Sun2017,Xiang2017}, magnetic order in our composition is expected to be stabilized
only at high pressures exceeding $40$\,kbar  \cite{Matsuura2017}, and a reduction of the superconducting fraction
was detected well-beyond $30$\,kbar in FeSe$_{0.88}$S$_{0.12}$ \cite{Yip2017}.

To elucidate the nature of quasi-particle scattering in the proximity of the nematic quantum phase transition, we investigate the resistivity exponent $n$, defined as $\rho(T)=\rho_0+ AT^{n}$, as a function of pressure and temperature.
 For a non- or weakly interacting Fermi liquid, one expects $n = 2$.
  In contrast, for systems in the proximity of the quantum critical points,
  enhanced order parameter fluctuations can lead to non-Fermi liquid behaviour with an exponent $n<2$  and additional scattering \cite{Rosch1999,Maslov2011,Barber2018,Lederer2017}.
  The precise power law $n$ value is influenced  by the nature of the critical fluctuations (their wave-vector and dynamical exponent),   by dimensionality \cite{Lohneysen2007} by the amount of disorder present \cite{Rosch1999}
  and by nearby quantum critical points \cite{Oliver2017}. 
     At  low temperatures away form the critical points in a clean system, 
   Fermi-liquid like behaviour is often recovered \cite{Lohneysen2007}.
   In this regime, the $A$ coefficient,  which is a measure of the strength of the electronic correlations  as $A \sim (m^*/m_e)^2$,    diverges upon approaching the critical region as  fluctuations become critical 
   ($m^*$ is the quasiparticle effective mass)
   \cite{Kasahara2010,Analytis2014}.

 Figure ~\ref{fig1}E shows the temperature and pressure dependence of $n$ as a color map (defined as $n = \partial \log(\rho - \rho_0) / \partial \log T)$ with $\rho_0$ extrapolated from low temperatures) from which several different non-Fermi liquid scattering regimes can be identified.
Firstly, inside the nematic phase, the resistivity follows a sublinear dependence, with $n$ ranging from $\sim 0.4 - 1$, similar to previous observations as a function of chemical pressure \cite{Reiss2017,Bristow2018,Urata2016} or in FeSe under applied pressure \cite{Knoner2015}.
Secondly, in the tetragonal phase
surrounding the nematic phase above $p_c$,
the resistivity displays a dominant $n \sim 3/2$ dependence over a wide temperature and pressure range (Figs.~\ref{fig1}B and C);
small deviations towards a $ n = 4/3$ power law are found for intermediate temperatures ($T <50$ K) for pressures close to $p_c$ (Fig.~\ref{fig1}B).
Thirdly, in the high-pressure region (close to $20$\,kbar), the resistivity has an extended region of almost linear temperature dependence (Fig.~\ref{fig1}A), similar to related pressure studies in the vicinity of the maximum $T_c$ for this composition \cite{Matsuura2017}.
The observation of an (almost) linear dependence of the resistivity is unusual and is often associated with antiferromagnetic fluctuations close to a magnetic quantum critical point which, in this case, is away from the expected nematic critical point around $p_c$ \cite{Kasahara2010}.

Next, we comment on the possible origin of the non-Fermi liquid behaviour outside the nematic phase, where critical
fluctuations are expected to affect scattering and the relevant resistivity exponents.
Above $p_c$,  a prominent exponent $n = 3/2$ can be assigned to strong antiferromagnetic fluctuations in the dirty limit \cite{Rosch1999}.
However, an NMR study under pressure on a compound with similar sulfur concentration ($x \sim 0.12$) suggested that
spin fluctuations are suppressed outside the nematic phase above $p_c$ and a regime
of nematic fluctuations extends towards $\sim 20$\,kbar \cite{Kuwayama2018}.
This would imply that the $n \sim 3/2$ region of non-Fermi liquid behaviour in resistivity can be caused by a region of strong nematic fluctuations.
Theoretically, the nematic critical fluctuations  can be
exceedingly effective in destroying quasiparticles and can
produce a striking nodal-antinodal dichotomy
generating  non-Fermi liquid-like behaviour over much of the Fermi surface enhancing resistivity
\cite{Lederer2017}.
Furthermore,  the proximity to a nematic quantum critical behaviour in certain
quasi-2D systems was associated to a resistivity exponent of $n = 4/3$ \cite{Maslov2011}
and the nematoelastic coupling with the lattice can also generate non-Fermi liquid-like behaviour
due to the presence of robust hot spots on the electron pockets
 \cite{Paul2017}.

Near the critical pressure, $p_c$, the resistivity in our single crystal of FeSe$_{0.89}$S$_{0.11}$ tuned by applied pressure does not follow  a linear $T$ dependence, as found near magnetic critical points \cite{Kasahara2010}
nor does it display the expected quantum critical fan with a constant exponent \cite{Lohneysen2007}.
Instead, the nematic critical region is
 best represented by a temperature-dependent resistivity exponent $n$, shown in Fig.\ref{fig1}B,
that evolves from a value around $n=3/2$ at high temperatures towards the $n=2$ value in the low temperature regime.
In fact, $n$ shows a marked upturn below 30~K for pressures close to $p_c$ that can be linearly extrapolate towards $n=2$,
as shown by dashed lines in Fig. \ref{fig1}B.
To further demonstrate this result, we can use magnetic fields to suppress superconductivity,
and by performing a two-band model fit to the magnetotransport data \cite{Proust2016}, we can
extrapolate the zero-field resistivity in the absence of superconductivity (Fig. S2).
At the lowest temperatures, we find a Fermi-liquid behaviour for all pressures across $p_c$, as shown in Fig.~\ref{fig1}D
(see also Figs. S1 and S3). Furthermore, the corresponding $A$ coefficient as well as the zero-temperature resistivity, $\rho_0$, are continuously suppressed with increasing pressure across the transition
(Figs.~\ref{fig4}B and S1). This strongly suggests that the nematic fluctuations do not become critical at $p_c$.

\begin{figure*}[htbp]
	\includegraphics[trim={0cm 0cm 0cm 0cm}, width=1\linewidth,clip=true]{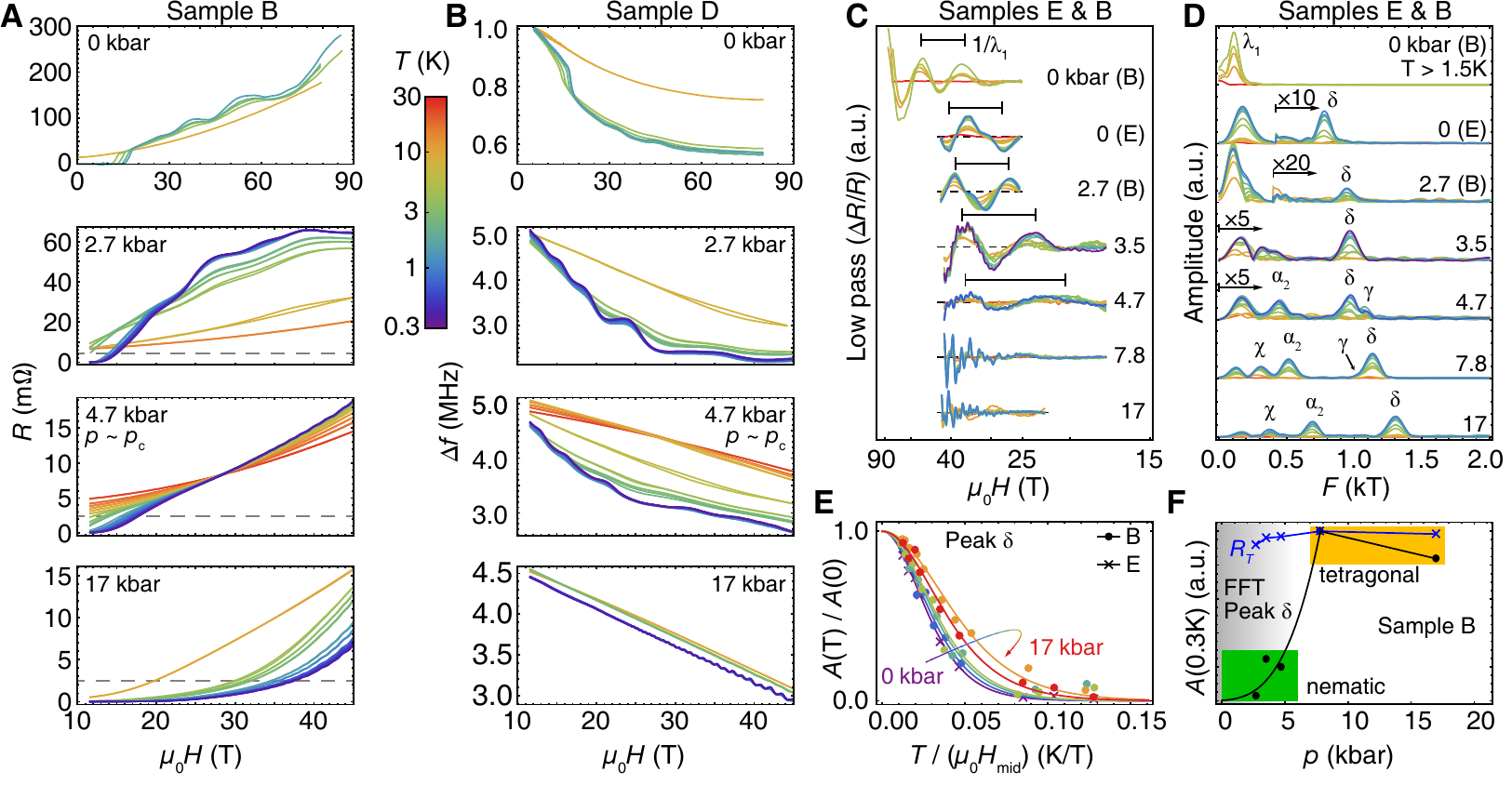}
	\caption{{\bf Evolution of quantum oscillations with pressure in FeSe$_{0.89}$S$_{0.11}$. }
		(\textbf{A}) Magnetotransport and (\textbf{B}) TDO resonant frequency variation, $\Delta f$ (MHz), for several different samples as a function of applied pressures (see also Figs.~S4-S5). Horizontal dashed lines in \textbf{A} indicate the sample resistance in zero field at the onset of superconductivity. (\textbf{C}) The oscillatory component of the magnetoresistance (after subtracting of a smooth background with a low-pass filter applied) shows the evolution of the dominant low-frequency oscillation. Dashed lines here indicate the base line.
		The horizontal line indicates the period of the dominant low frequency ($\lambda_1$). (\textbf{D}) Fourier transform of the oscillatory part of the magnetotransport data in (A) after the background subtraction. The distinct peaks identified correspond to extremal cross-sections of the Fermi surface. (\textbf{E}) The temperature dependence of the amplitude gives the cyclotron effective band mass of the $\delta$ orbit for different constant pressures. Lines are fits to the Lifshitz-Kosevich mass damping term \cite{Shoenberg1984}. (\textbf{F}) Low-temperature amplitude variation with pressure of the $\delta$ orbit (solid symbols) and the expected amplitude variation due to the mass damping term $R_T$ (cross symbols). Solid lines are guides to the eye. Data in panels \textbf{C} and \textbf{D} are shifted vertically for clarity.}
	\label{fig2}
\end{figure*}

{\bf The evolution of the Fermi surface with pressure.}
Next we use quantum oscillations measurements under pressure
to follow the evolution of the Fermi surface and to assess the strength of electronic correlations across the phase diagram of FeSe$_{0.89}$S$_{0.11}$ under pressure.
Figures~\ref{fig2}A and B show the field dependence of the resistance and TDO data for pressures up to $17$\,kbar and up to $45$\,T, and the ambient pressure
measurements up to $80$\,T for three different samples. Results on additional samples and pressures are shown in Figs. S4-S8.
There are different field regimes seen in the raw data:
a) the superconducting state with zero resistance at low magnetic fields, b) finite resistance that increases
strongly with magnetic field in the crossover vortex-liquid region to the normal state at higher fields
and c) normal magnetoresistance accompanied by quantum oscillations in high magnetic fields.

At low pressures in the nematic state, the magnetotransport data
is dominated by a low-frequency oscillation (Fig.~\ref{fig2}A). With increasing
pressure, the frequency is reduced and the oscillation
disappears beyond  $4.7$\,kbar, in the proximity of $p_c$
(Figs.~\ref{fig2}A, B and C).
This behaviour is remarkably similar to the Lifshitz transition observed
on the border of the nematic phase transition,
also identified in quantum oscillations as a function of chemical pressure in FeSe$_{1-x}$S$_x$ \cite{Coldea2016}.
At high pressures and in high fields, a high-frequency oscillation is visible in the raw data (see Figs.~\ref{fig2}A and B), associated with a large Fermi surface sheet.
This rules out that a reconstruction of the Fermi surface occurs at high pressures in this compound, similar to
to the tetragonal phases of FeSe$_{1-x}$S$_x$ \cite{Coldea2016}, but in contrast to the magnetic phase in FeSe under pressure \cite{Terashima2016}.

\begin{figure}[htbp]
	\centering
	\includegraphics[trim={0cm 0cm 0cm 0cm}, width=0.9\linewidth,clip=true]{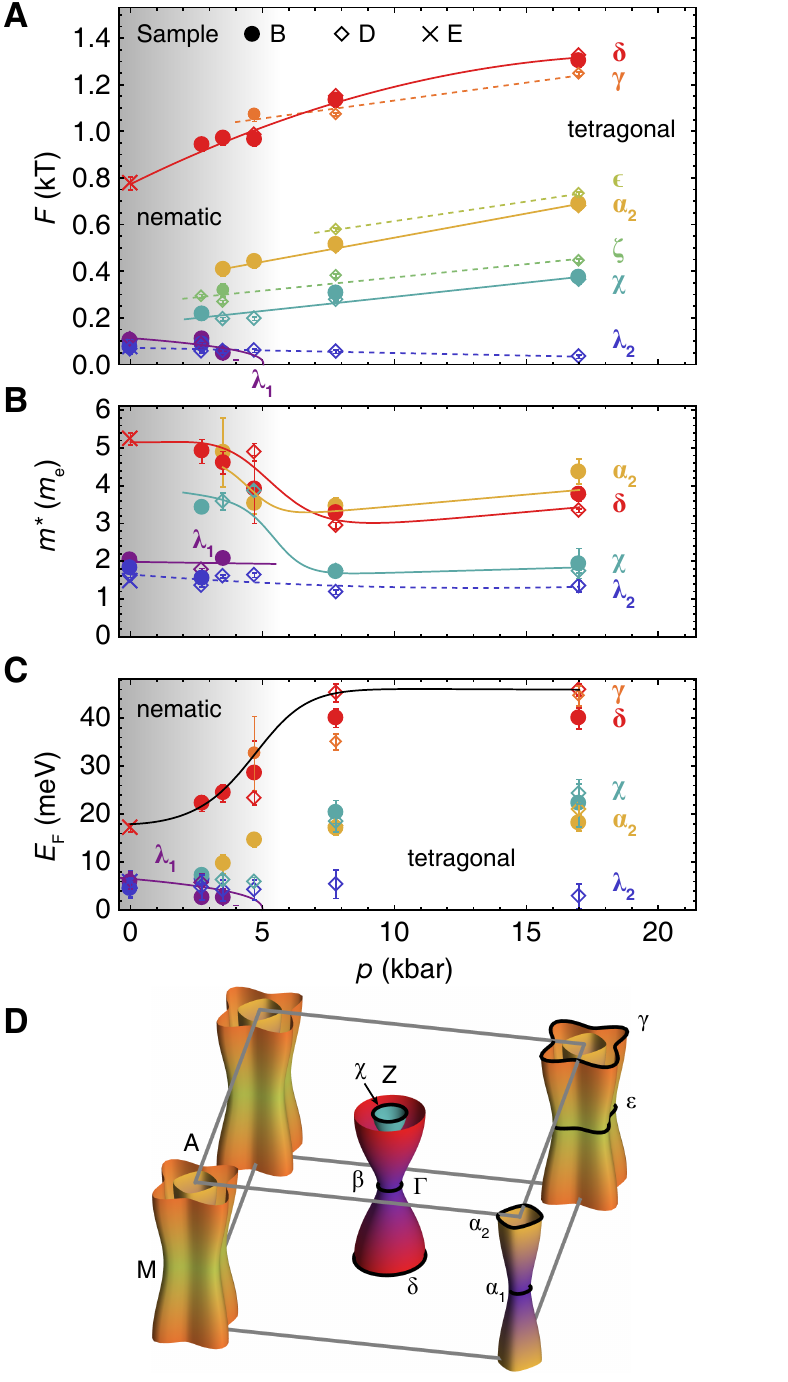}
	\caption{{\bf Pressure tuning of Fermi surface and electronic correlations.}
				(\textbf{A}) Quantum oscillation frequencies, and
		(\textbf{B}) the cyclotron masses as a function of pressure for samples B, D and E.
The masses associated to
 $\chi$ and $\delta$ orbits
 show a clear decrease upon crossing the nematic phase transition.
A topological Lifshitz transition
occurs at $p_c$ and one small orbit in the $k_z$=0 plane disappears (either $\beta$ or $\alpha_1$ can be assigned to $\lambda_1$). (\textbf{C}) The Fermi energy $E_F$ assuming parabolic band dispersions, based on Fermi surface cross-sections and quasiparticle masses (panels A and B). Lines are guides to the eye. The splitting of the $\gamma$ and $\delta$ frequencies is summarized in Fig.~S7.  Shaded areas indicate the nematic phase
(\textbf{D})  Sketch of the Fermi surface in the nematic phase and the different two-dimensional
orbits, based on ARPES and quantum oscillation measurements \cite{Coldea2016,Reiss2017,Watson2015c}.}
	\label{fig3}
\end{figure}

\begin{figure}[htbp]
	\centering
	\includegraphics[trim={0cm 0cm 0cm 0cm}, width=8cm,clip=true]{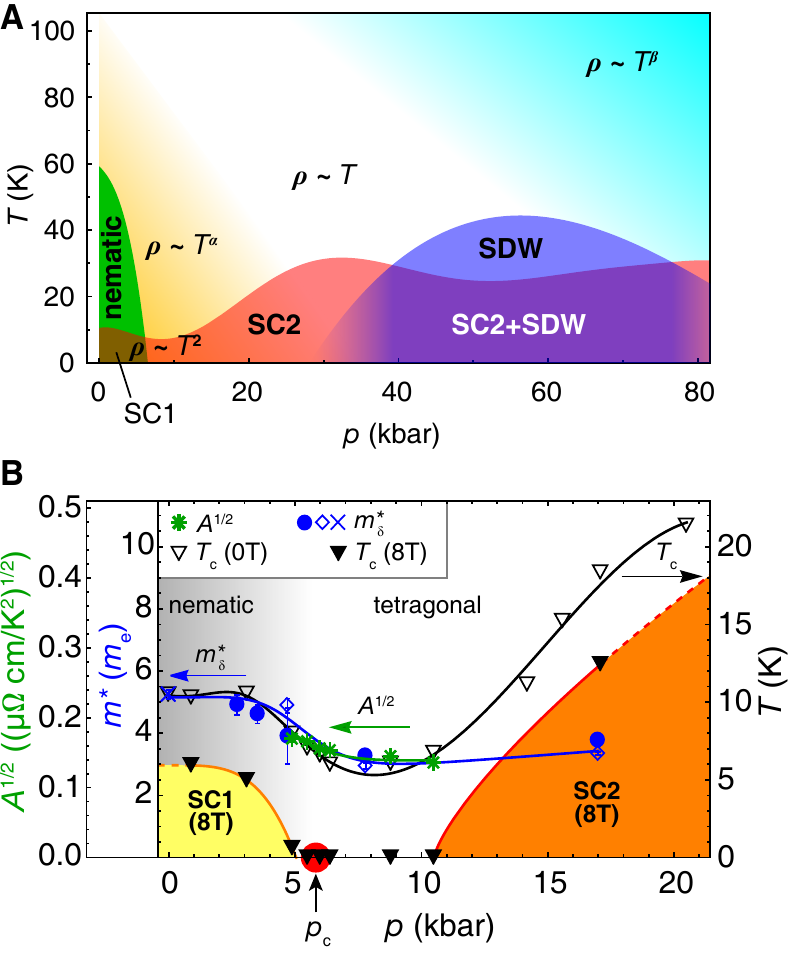}
	\caption{{\bf Pressure-temperature phase diagrams.}
		(\textbf{A}) Extended phase diagram of FeSe$_{0.89}$S$_{0.11}$ based on our work and previous reports \cite{Matsuura2017}.
		(\textbf{B}) Comparison of the band masses of the $\delta$ orbit with $T_c$ and the Fermi liquid coefficient, $A$.
A striking similar pressure dependence of electronic correlations and superconductivity is maintained across the nematic phase transition,
but it is lost towards the high pressure high-$T_c$ state.
The temperature-pressure phase diagram in a static magnetic field of $8$\,T shows the separation of the two superconducting domes unmasking the nematic quantum phase transition at $p_c$.
Symbols for the masses of the $\delta$ orbits are identical to Fig.~\ref{fig3}. Solid lines are guides to the eye and dashed lines are extrapolations.}
	\label{fig4}
\end{figure}

To quantify the complex oscillatory spectra of quantum oscillations,
we use both a fast Fourier transform  in Fig.~\ref{fig2}D
(after removing a smooth and monotonic polynomial background from the raw data)
or directly fitting the Lifshitz-Kosevich formalism to the raw data in field (with a low-pass filter applied),
as shown in Figs.~\ref{fig2}C, S4 and S6.
This analysis reveals several distinct peaks at ambient pressure, consistent with a complex multi-band electronic structure.
Based on ARPES data and previous quantum oscillation measurements,
the ambient pressure Fermi surface is formed of two concentric electron-like
 and one outer hole-like quasi-two dimensional sheets ($\beta$ and $\delta$ orbits) as well as a small inner 3D hole pocket centered at the Z-point  ($\chi$),
 shown in Fig.\ref{fig3}D \cite{Watson2015c,Coldea2017,Reiss2017}.
  As a function of applied pressure, the cylindrical Fermi surfaces can become warped
  along the $k_z$ direction, either due to changes in the degree of interlayer hopping term  and/or electronic correlations  leading to a Lifshitz transition and the disappearance of the neck orbits (like $\beta$ or $\alpha_1$ in Fig.\ref{fig3}D).
 The multi-band structure of  FeSe$_{0.89}$S$_{0.11}$ could give rise up to a maximum of seven frequencies depending on pressure assuming
 that all the cross-sectional areas are dissimilar for different Fermi surface sheets.
 In our quantum oscillations data, we can distinguish fewer frequencies but we observe clearly
the high frequency region, dominated by the largest orbit of the hole band, $\delta$, as shown in  Fig.~\ref{fig2}D.
Weaker features associated to the largest orbit of the electron band, $\gamma$, can be also detected at some pressures
and in the TDO signal, as shown in Fig.~S7.

The dominant low frequency, $\lambda_1$, can be extracted by directly fitting of the data in Fig.~S6,
which confirms that this low frequency disappears close to $p_c$ (see $\lambda_1$ in Fig.\ref{fig3}).
Since this observation is in agreement with 
previous studies using sulfur substitution as the tuning parameter,
we attribute this  Lifshitz transition, caused by the change
in topology of the Fermi surface, as a common feature of the electronic
structure at the nematic phase transition tuned either by chemical or applied pressures.
Furthermore, the TDO data over the entire available magnetic field range
seem to indicate the presence of an additional 
low-frequency oscillation beyond
the critical nematic phase (see Figs.~S4 and S6).

With increasing pressure all frequencies increase in size, with the exception of the lowest two frequencies,
as summarized in Fig.~\ref{fig3}A.
At the highest measured pressure, the Fermi surface area of the $\delta$ pocket, $A_{k}$,
given by the Onsager relation \cite{Shoenberg1984}, $F = A_{k} \hbar /(2 \pi e)$, has expanded almost by a factor two,
 as compared with its size at the ambient pressure. This is much more than the expected growth of the Brillouin zone
 (less than 4\%), assuming a simple contraction of the in-plane unit cell \cite{Kothapalli2016,Tomita2015}.
This implies significant pressure-induced topological changes of the Fermi surface,
similar to those observed as a function of chemical pressure \cite{Coldea2016}.

{\bf Quasiparticle masses.}
We now turn to the evolution of the
electronic correlations across the phase diagram.
The evolution of the cyclotron effective mass, $m^*$,
is extracted from the temperature dependence of the quantum oscillation amplitudes,
analyzed quantitatively within
the Lifshitz-Kosevich formalism (Fig.~\ref{fig2}E) \cite{Shoenberg1984},
and summarized in Fig.~\ref{fig3}B.
The cyclotron mass of the highest frequency, $\delta$,
(as well as $\chi$ and $\gamma$ in Fig.~S7),  decreases significantly
with increasing pressure from $5$\,$m_e$ to $3$\,$m_e$,
signifying a strong reduction in electronic correlations away from the nematic phase transition.
This shows that the quasiparticle mass divergence is avoided
in the vicinity of the nematic quantum phase transition around $p_c$.
We find similar trends in the evolution of the
$A$ coefficient and the quasiparticle mass of the $\delta$ orbit  ($m^*_{\delta}$)  
across the nematic quantum phase transition, as shown in Fig.~\ref{fig4}B. 
In the high pressure regime, a slight increase in the effective masses
of the $\delta$ and $\xi$ orbits is observed, but their
values remain smaller than in the nematic phase, despite
the fact that superconductivity has increased twice.

{\bf Quantum oscillation amplitude variation.}
An unusual observation of our study
is the significant variation of the quantum oscillations amplitude of the highest frequency $\delta$ across
the nematic phase transition, shown in Fig.~\ref{fig2}F.
According to the Lifshitz-Kosevich formalism, the amplitude of quantum oscillations
is affected by different damping terms directly linked to changes of the quasiparticle masses,
impurity scattering, spin-splitting effects, as well as the curvature factor of the Fermi surface \cite{Shoenberg1984}.
As the same single crystal is pressurized, the level of impurity scattering
is not expected to change over the entire
pressure range. This is supported by the Dingle analysis that
shows a similar impurity scattering rate at all pressures in Fig.~S8.
Furthermore, spin-splitting effects are likely to be negligible for large Fermi surface sheets in a non-magnetic system \cite{Shoenberg1984}.
As the amplitude variation strongly exceeds the
expectation linked to the changes in the quasiparticle masses, as shown in Fig.~\ref{fig2}F,
other effects must play an important role.
The curvature factor is the largest for a two-dimensional Fermi surface \cite{Shoenberg1984}
but at the Lifshitz transition an increase in the warping and three-dimensionality of the Fermi surface with pressure
would generate a smaller curvature factor \cite{Mandal2014}.
Lastly, NMR studies have revealed strong antiferromagnetic fluctuations within the nematic phase,
before being suppressed close to the nematic phase transition \cite{Kuwayama2018}.
Based on this, the unusual suppression of the amplitude of the
high frequency quantum oscillation, $\delta$, within the nematic phase
may be linked to smearing of the Fermi surface caused by the
strength of the antiferromagnetic fluctuations within the nematic phase.
Resistivity data within the nematic phase has an almost linear dependence (see Fig.\ref{fig1})
above $T_c$ suggesting that antiferromagnetic fluctuations may be relevant
deep inside the nematic phase. At the lowest temperatures below $p_c$,
a resistivity exponent $n$ is harder to estimate reliably
due to the larger upper critical fields and the drastic changes in magnetoresistance
caused by the low frequency quantum oscillations (see Fig.S1).

{\bf Discussion.} The size of the Fermi energy, $E_{\rm F}$, in a multiband
superconductor is a parameter that provides important clues
about the stabilization of different ground states competing with superconductivity \cite{Chubukov2016}.
In the case of a small Fermi energy, a Pomeranchuck instability
is favoured, whereas a spin-density wave is
stabilized for large Fermi energies.
From our quantum oscillation study, we find that the largest Fermi energies increase significantly
outside the nematic phase as a function of pressure in FeSe$_{0.89}$S$_{0.11}$, as shown in Fig.~\ref{fig3}C.
Large hole and electron Fermi surface sheets at high pressures are expected to enhance nesting,
promoting the stabilization of a magnetically ordered-state, such as the spin-density wave (SDW in Fig.~\ref{fig4}A),
identified by other pressure studies  \cite{Matsuura2017}.
The small pockets in Fig.~\ref{fig3}C have Fermi energies of the order 3-5 meV,
comparable to the superconducting gap, $\Delta$ (Fig.S4);
under such conditions a BCS-BEC crossover can partially occur \cite{Kasahara2014}.

Based on our experimental findings,  the superconducting phase
of FeSe$_{0.89}$S$_{0.11}$ under pressure 
is composed of two different superconducting domes, separated by a Lifshitz transition.
The first one at low pressure emerges from the nematic phase (SC1 in Fig.\ref{fig4}B) and
 the second one at high pressure approaching the spin-density wave phase (SC2 in Figs.\ref{fig4}A, B).
The two domes can also be visualized in the pressure-temperature phase diagram in magnetic fields of 8~T in Fig.\ref{fig4}B.
Remarkably, the nematic quantum
phase transition is located in the normal state between
the two superconducting domes.

The cyclotron effective masses measure that the strength of electronic correlations varies with pressure,
being in general larger inside the nematic phase compared with those in the tetragonal phase.
Interestingly, we find a quantitative link between the cyclotron masses of the $\delta$ orbit,
the $A$ coefficient from the low temperature resistivity data, and the value of $T_c$ in the nematic phase and in the low-$T_c$ tetragonal phase. This supports that
the hole bands are closely involved in the pairing mechanism (Fig.~\ref{fig4}B).
This correlation between the band renormalisation and $T_c$ was also captured by quantum oscillations experiments
and ARPES studies on FeSe$_{1-x}$S$_{x}$ \cite{Coldea2016,Reiss2017}.
Furthermore, NMR studies show that the strength of antiferromagnetic fluctuations also correlates with $T_c$ within the nematic phase,
being suppressed  together with the nematicity \cite{Kuwayama2018,Wiecki2018}.
For higher pressures, the correlation between the hole-like band masses
and $T_c$ is lost, suggesting changes in the pairing mechanism towards the high-$T_c$ phase.

Upon approaching a critical region as a function of the tuning parameter, one would expect the divergence of the quasiparticle
masses, $A$ coefficient and an enhanced superconducting phase
  \cite{Shishido2010,Analytis2014,Lohneysen2007}.
Here, near the nematic critical point at $p_c$, in the absence of magnetic order,
we find a smoothly evolving $A$ coefficient and non-divergent changes in the effective masses for the large, well-defined orbit ($\delta$).
This suggests that the nematic fluctuations are finite and not critical at $p_c$.
Strikingly, the superconducting transition temperatures are minimal
in the vicinity of the nematic quantum  phase transition (Fig.~\ref{fig4}).
 Therefore, superconductivity is not enhanced by non-critical  nematic fluctuations in the absence of  the magnetic
order. 
This is in contrast to other iron-based superconductors 
where critical nematic and/or magnetic fluctuations enhance superconductivity
approaching a spin-density wave state 
\cite{Kasahara2010,Lederer2017}.
Based on these observations,
we interpret our results  in the context of quenched nematic criticality due to a strong
 nematoelastic coupling to the lattice  \cite{Paul2017,Labat2017}.
   This coupling is responsible for the suppressed superconductivity
 \cite{Labat2017} and the finite quasiparticle mass
 in the proximity of the nematic quantum critical point
   (compared with data in Fig.~S7).

The transport across the two superconducting domes tuned by pressure
is dominated by different types of scattering. These processes can be either
completely independent of each other or they can overlap in certain pressure regions.
A linear resistivity is observed over a large temperature range in the high pressure regime
in the vicinity  of a antiferromagnetic quantum critical point, expected around $30$\,kbar  \cite{Matsuura2017}.
On the other hand, outside nematic phase boundaries, we find
a non-Fermi liquid behaviour, with a dominant $T^{3/2}$ resistivity  surrounding the nematic phase,
consistent with the region dominated by strong nematic fluctuations, as identified by an NMR study \cite{Kuwayama2018}.
  In reality, the resistivity exponent $n$ close to $p_c$ in our study 
is not constant and it has a strong temperature dependence (Figs.~\ref{fig1}B).
The exponent changes from  $n = 3/2$ at high temperatures towards 
the Fermi liquid regime 
with $n = 2$  at low temperature  
for all pressures across the nematic quantum phase transition.  
This is consistent with the predictions of the model
assuming a strong nematoelastic coupling that 
can generate non-Fermi liquid behaviour at high temperatures
and the transformation into a Fermi liquid below $T_{FL}$.
We can estimated that for our system  $T_{FL}\sim 13 $~K 
 (from ARPES and nematic susceptibility measurements for
similar compositions \cite{Watson2015c,Hosoi2016}), 
in good agreement with our experiments (Fig.\ref{fig1}).
Thus, the transport behaviour near the nematic critical point revealed by our study is
 very different from the linear $T$-behaviour in the proximity of a magnetic critical point  \cite{Kasahara2010}.

  The resistivity behaviour in the vicinity  of the nematic critical point tuned using applied hydrostatic pressure seems to be
  in contrast to studies using chemical pressure in FeSe$_{1-x}$S$_x$,
  which identified a linear-$T$ region at the nematic quantum phase transition and
  equally no evidence of an antiferromagnetic phase for the highest available doping was found \cite{Urata2016,Licciardello2019}.
    While in our study a single high quality single crystal is tuned as a function of pressure across different phases,
   studies using chemical substitution require different samples to be measured and the varying impurity scattering between samples
 needs to be taken into account.
   In fact, the role played by impurities on quantum critical behaviour can have drastic effects on scattering.
    For example, critical antiferromagnetic fluctuations cause strong scattering only in the vicinity of hot-spots on the Fermi surface which can be short-circuited by other parts of the Fermi surface in the ultra-clean limit giving a Fermi-liquid $T^2$ resistivity \cite{Lohneysen2007,Oliver2017}.
    Impurities or  multiple scattering modes can make the scattering more isotropic and could produce a $T$-linear resistivity close to an 
    antiferromagnetic critical point in a two-dimensional weakly disorder metal  
     \cite{Rosch1999}.
     Similar arguments can be applied to nematic critical systems
     where nematoelastic coupling generate isolated hot spots on the Fermi surface \cite{Paul2017,Lederer2017,Wang2019}. 

Remarkably, the fine tuning using hydrostatic pressure
in FeSe$_{0.89}$S$_{0.11}$ provides an unique insight into the behaviour of a pristine nematic quantum critical point.
The electronic signature of this point is substantially different from other critical points
in iron-based superconductors, where both magnetic and nematic phases
closely coexist. Superconductivity is not enhanced by non-critical nematic fluctuations, in the absence of magnetic order,
as they are cut off along certain directions by the noncritical shear strains
that are invariably present in real materials \cite{Labat2017}.
Our findings should have profound implications on
understanding the superconducting
mechanisms  in different classes of superconductors where nematic phases are present.

\vspace{0.5cm}
{\bf Acknowledgements}
We thank Penglin Cai for technical support provided with setting up the PPMS pressure cell
and Andrey  Chubukov, Erez Berg, Raphael Fernandes, Indranil Paul, Roser Valenti, Ilya Vekhter, Matthew Watson,
Zachary Zajicek  for useful discussions and comments.
This work was mainly supported by EPSRC (EP/I004475/1, EP/I017836/1).
AAH acknowledges the financial support of the Oxford Quantum
Materials Platform Grant (EP/M020517/1).
A portion of this work was performed at the National High Magnetic Field Laboratory, which is supported by National Science Foundation Cooperative Agreement No. DMR-1157490 and the State of Florida.
Part of this work was supported by
HFML-RU/FOM and LNCMI-CNRS, members of the European Magnetic Field
Laboratory (EMFL) and by EPSRC (UK) via its membership to the EMFL
(grant no. EP/N01085X/1).
Part of this work at the LNCMI was supported by Programme
Investissements d'Avenir under the program ANR-11-IDEX-0002-02, reference ANR-10-LABX-0037-NEXT.
AIC thanks the hospitality
of KITP supported by the National Science Foundation under
Grant No. NSF PHY- 1125915.
We also acknowledge financial support of the John Fell
Fund of the Oxford University.
AIC acknowledges an EPSRC Career Acceleration Fellowship (EP/I004475/1).

\vspace{0.5cm}
{\bf Methods}\\
{\bf \small Sample characterization.}
Single crystals of FeSe$_{0.89}$S$_{0.11}$ were grown by the KCl/AlCl$_3$ chemical vapour transport method, as reported previously \cite{Bohmer2013,Bohmer2016g}. More than 20 crystals were screened at ambient pressure and they showed sharp superconducting transition regions of $\sim 0.1$\,K, and large residual resistivity ratios up to 17 between room temperature and the onset of superconductivity. Crystals from the same batch were previously used in quantum oscillations and ARPES studies \cite{Watson2015a,Coldea2016}.

{\bf \small Magnetotransport measurements.}
High magnetic field measurements up to 45\,T at ambient pressure and under hydrostatic pressure were performed using the hybrid magnet dc facility at the NHMFL in Tallahassee, FL, USA. Pressures up to 17\,kbar were generated using a NiCoCr piston cylinder cell, using Daphne Oil 7575 as pressure medium. The pressure inside the cell was determined by means of ruby fluorescence at low temperatures where quantum oscillations were observed. Different samples were measured simultaneously under pressure: sample B was used for transport measurements, and sample D-TDO was positioned inside a coil and the resonant frequency of an LC tank circuit driven by a tunnel diode was recorded (TDO). Transport sample E was measured at ambient pressure at the same time. Magnetotransport measurements were performed using the standard $ac$ technique.
Pulsed-field measurements up to 80\,T at ambient pressure were carried out in Toulouse, using the same samples (transport measurements on sample B, TDO measurements on sample D).
Magnetotransport and Hall effect measurements under pressure using a 5-contact configuration were carried out on sample A in low fields up to 16\,T in an Oxford Quantum Design PPMS and an ElectroLab High Pressure Cell, using Daphne Oil 7373 which ensures hydrostatic conditions up to about 23\,kbar. The pressure inside this cell was determined via the superconducting transition temperature of Sn after cancelling the remanent field in the magnet.
The magnetic field was applied along the crystallographic $c$ axis for all samples.
A maximum current of up to 2\,mA flowing in the conducting tetragonal $ab$ plane was used.

\bibliography{FeSeS_bib_pressure_feb19}

\end{document}